\documentclass[useAMS,usenatbib]{mn2e}

\usepackage{graphicx}
\usepackage{psfig}
\usepackage{amsmath}
\usepackage{multirow}
\usepackage{booktabs}

\newcommand{\lrmv}{\ell_{\rm rmv}}
\newcommand{\WMAP}{\textit{WMAP}}
\newcommand{\vecn}{\textbf{\textit{n}}}
\newcommand{\vecp}{\textbf{\textit{p}}}
\newcommand{\veck}{\textbf{\textit{k}}}

\title[Comparison of WMAP5 Local Extrema with Two Anisotropic Models]
{Frequentist comparison of CMB local extrema statistics in the five-year \WMAP\ data
with two anisotropic cosmological models}

\author[Hou et al.]
{Zhen Hou$^{1,2,3}\footnote{E-mail:houzhen@pmo.ac.cn}$,
A.\,J. Banday$^{4,2}$, K.\,M. G\'orski$^{5,6,7}$,  N.\,E. Groeneboom$^{8}$, \and H.\,K. Eriksen$^{8,9}$ \\
$^{1}$ Purple Mountain Observatory, Chinese Academy of Sciences, 210008, Nanjing, China\\
$^{2}$ Max-Planck-Institute for Astrophysics,
Karl-Schwarzschildstrasse 1, D-85741, Garching bei M\"{u}nchen, Germany\\
$^{3}$ Graduate University of Chinese Academy of Sciences, 100049,
Beijing, China\\
$^{4}$ Centre d'Etude Spatiale des Rayonnements, 9 av du Colonel Roche, BP 44346, 31028 Toulouse Cedex 4, France\\
$^{5}$ Jet Propulsion Laboratory, 4800 Oak Grove Drive, Pasadena CA 91109, USA\\
$^{6}$ California Institute of Technology, Pasadena, CA 91125, USA\\
$^{7}$ Warsaw University Observatory, Aleje Ujazdowskie 4, 00-478 Warszawa, Poland\\
$^{8}$ Institute of Theoretical Astrophysics, University of Oslo,
P.O. Box 1029 Blindern, N-0315 Oslo, Norway \\
$^{9}$ Centre of Mathematics for Applications, University of Oslo,
P.O. Box 1053 Blindern, N-0316 Oslo, Norway \\
}

\pagerange{\pageref{firstpage}--\pageref{lastpage}}

\pubyear{2009}

\begin{document}

\maketitle

\label{firstpage}

\begin{abstract}
We present local extrema studies of two models that introduce a
preferred direction into the observed cosmic microwave background
(CMB) temperature field.  In particular, we make a frequentist
comparison of the one- and two-point statistics for the dipole
modulation and ACW models with data from the five-year
\textit{Wilkinson Microwave Anisotropy Probe} (\WMAP). This analysis
is motivated by previously revealed anomalies in the \WMAP\ data,
and particularly the difference in the statistical nature of the
temperature anisotropies when analysed in hemispherical partitions.

The analysis of the one-point statistics indicates that the
previously determined hemispherical variance difficulties can be
apparently overcome by a dipole modulation field, but new
inconsistencies arise if the mean and the $\ell$-dependence of the
statistics are considered.  The two-point correlation functions of
the local extrema, $\xi_{\rm TT}$ (the temperature pair product) and
$\xi_{\rm PP}$ (point-point spatial pair-count), demonstrate that
the impact of such a modulation is to over-`asymmetrise' the
temperature field on smaller scales than the wave-length of the
dipole or quadrupole, and this is disfavored by the observed data.
The results from the ACW model predictions, however, are consistent
with the standard isotropic hypothesis. The two-point analysis
confirms that the impact of this type of violation of isotropy on
the temperature extrema statistics is relatively weak.

From this work, we conclude that a model with more spatial structure
than the dipole modulated or rotational-invariance breaking models are
required to fully explain the observed large-scale anomalies in the \WMAP\
data.
\end{abstract}

\begin{keywords}
methods: data analysis -- cosmic microwave background.
\end{keywords}
\footnotetext{E-mail: houzhen@pmo.ac.cn}

\section{Introduction}
\label{sec_intro}

Local extrema in the CMB have been extensively studied in the
context of hotspots (peaks) and coldspots (troughs) arising in
Gaussian random fields \citep{bond_etal_1987, vittorio_etal_1987}.
Additional statistics related to such local extrema, such as the
Gaussian curvature and temperature-correlation function, have been
investigated as a means to distinguish the geometry of the universe
and test the Gaussian hypothesis for the nature of the initial
conditions \citep{barreiro_etal_1997, heavens_etal_1999,
heavens_etal_2001}. Such statistical techniques have subsequently
been applied to several datasets, and most notably by
\citet{kogut_etal_1995, kogut_etal_1996} with the $COBE$-DMR data.
Observations from the \textit{Wilkinson Microwave Anisotropy Probe}
(\WMAP) currently provide the most comprehensive, full-sky,
high-resolution CMB measurements to-date, and studies on the local
extrema properties thereof have been undertaken \citep{LW04, LW05,
  tojeiro_etal_2006}.

More recently in \citet{hou_etal_2009}, we extensively analyzed the
statistical properties of both the one- and two-point statistics of
local extrema in the five-year \WMAP\ temperature data. Such extrema
are defined as those pixels whose temperature values are higher
(maxima) or lower (minima) than all of the adjacent pixels
\citep{wandelt_etal_1998}. We considered only that part of the sky
outside of the \WMAP\ KQ75 mask and its subsequent partition in
either Galactic or Ecliptic coordinates into northern and southern
hemispheres (hereafter GN, GS, EN, ES).  A frequentist comparison
with the predictions of a Gaussian isotropic cosmological model that
adopted the best-fit parameters from the \WMAP\ team was then made.
The hypothesis test indicated a low-variance of both local maxima
and minima in the Q-, V- and W-band data that was inconsistent with
the Gaussian hypothesis at the 95\% C.L. The two-point analysis
showed that the observed temperature pair product at a given
threshold $\nu$, $\xi_{\rm TT}(\theta,\nu > 1,2)$, indicates a
$3\sigma$ level `suppression' on GN and EN, whereas $\xi_{\rm
TT}(\theta<20^{\circ},\nu < \infty)$ is suppressed on the full-sky
and both northern hemisphere partitions.  The latter is also the
case for the point-point spatial pair-count function $\xi_{\rm
PP}(\theta<20^{\circ},\nu > 1,2)$. Intriguingly, the statistics
showed an $\ell$-dependence such that consistency with the Gaussian
hypothesis was achieved once the first 5 or 10 best-fitting
multipoles were subtracted, implying that the anomalies may be
connected to features of the large-scale multipoles.

The local extrema anomalies therefore provide further evidence of a
hemispherical asymmetry that was originally revealed using the power
spectrum \citep{hansen_etal_2004, eriksen_etal_2004,
  hansen_etal_2008}, the N-point correlation functions
\citep{eriksen_etal_2005} and Minkowski functionals
\citep{park_2004}. This can be interpreted as a violation of the
cosmological principle of isotropy.  Theoretically,
\citet{gordon_etal_2005} proposed a mechanism of spontaneous
isotropy breaking in which the long-wavelength modes of a mediating
field couple non-linearly to the CMB perturbations. These
fluctuations appear locally as a gradient and imprint a preferred
direction on the sky. An implementation of a multiplicative
modulation field of the intrinsic anisotropy with a single preferred
direction was fitted to the three-year \WMAP\ observations in
\citet{eriksen_etal_2007} . The so-called \lq dipole modulation
field' (dmf) was detected at a significant confidence level, and
confirmed at even higher significance in the five-year \WMAP\ data
by \citet{hoftuft_etal_2009}.

Another mechanism that violates rotational invariance was
proposed by \citet{acw_2007}. The ACW model picks out a preferred
direction during cosmological inflation to modify the power spectrum
of primordial perturbations, $P(k)$. Its imprint on the CMB
can then be described by the covariance matrix of spherical
harmonic coefficients, $\langle a_{\ell m}a^{*}_{\ell'm'}\rangle$.
\citet{groe_hke_2009} estimated the parameters of the ACW model in an
extended CMB Gibbs sampling framework from the five-year \WMAP\
data. The posterior distribution of the parameters indicated a
convincing detection of isotropy violation at $3.8\sigma$ significance in the
W-band data.

In this paper, we make a frequentist comparison of the dmf and ACW
models with the five-year \WMAP\ maps of temperature anisotropy
using a large number of Monte-Carlo simulations. In particular, we
consider the one- and two-point statistics of local extrema and
evaluate whether the \WMAP\ data are more consistent with these
models as opposed to the standard Gaussian cosmological models
against which various anomalies have been claimed. A rigorous
hypothesis test methodology is applied to establish the significance
at which the models impact the previous conclusions regarding \WMAP\
data and the violation of isotropy. In addition, the $\xi_{\rm TT}$
and $\xi_{\rm PP}$ correlation functions are utilised to further
confirm the findings of the one-point analysis and to study the
scale-dependence of the local extrema both in real and spherical
harmonic space.

It might be considered that the statistical significance of the local
extrema anomalies alone is insufficient to warrant an analysis of the
anisotropic models, in particular given the additional parameters
required to specify them. However, that they provide significantly
better fits to the \WMAP\ data than the isotropic model has been
demonstrated by independent Bayesian power spectrum analysis. It is
clearly then quite legitimate to consider the local extrema statistics
in terms of these improved models, and to determine whether the
observed local extrema are also more consistent with such cosmological
prescriptions.  In fact, we will find that there is little evidence
that the extrema anomalies are remedied by these models, and indeed in
one case we show that the anomalous behaviour becomes more
significant.

This paper is organised as follows. In
Section~\ref{subsec_wmap_data}, we present an overview of the \WMAP\
data and the instrumental properties that must be involved in
Gaussian simulations to enable an unbiased comparison with the real
data. The two models and the algorithms for simulating them are
briefly reviewed in Section~\ref{subsec_2model}.
Section~\ref{subsec_analysis} prescribes the technique used for
further data-processing and introduces the one- and two-point
analysis methods. The results are reported in
Section~\ref{sec_results} before we present our conclusions in
Section~\ref{sec_conclusion}.

\section{METHOD}
\label{sec_method}

Both the dipole modulation and ACW models provide mechanisms for the violation
of cosmological isotropy in the context of Gaussian random fields. In a frequentist formalism,
we construct simulations for each model, convolve with the
appropriate instrumental beam transfer functions, then add
noise according to the properties of the \WMAP\ detectors.
The statistical properties of the simulated data
are then compared with the observed ones to
verify the consistency of the models with the observed Universe.

\subsection{The \WMAP\ data}
\label{subsec_wmap_data}

The \WMAP\ instrument consists of 10 differencing assemblies (DAs)
covering frequencies from 23 to 94 GHz \citep{bennett_etal_2003}.
The V and W-band foreground-reduced sky maps are used in our
analysis, consistent with the data selection for the \WMAP\
five-year power spectrum estimation \citep{nolta_etal_2009}. The
maps are available in the
HEALPix\footnote{http://healpix.jpl.nasa.gov/} pixelisation scheme
with $N_{\rm side}=512$ from the LAMBDA
website\footnote{http://lambda.gsfc.nasa.gov/product/map/dr3/\\maps\_da\_forered\_r9\_iqu\_5yr\_get.cfm}.
We combine data from DAs covering the same frequency using uniform
weighting over the sky and equal weight per DA. This allows a simple
effective beam to be computed from the combination of beam transfer
functions ($b_{\ell}$) for the DAs involved. In fact, the \WMAP\
beams are not axisymmetric and \citet{wehus_etal_2009} have
attempted to assess the effect of the combination of the true
asymmetric beams with scanning strategy on the sky using
reconstructed sky maps. They report deviations from the official
$b_{\ell}$\footnote{http://lambda.gsfc.nasa.gov/product/map/dr3/beam\_info.cfm}
at a level of $\approx$1.0\% at $\ell\approx800$. However, the
additional smoothing employed in this work render our statistics
insensitive to such deviations, thus we continue to adopt the beam
properties provided by the \WMAP\ team. There are two noise
components contributing to observations, white noise and residual
$1/f$ noise following an effective destriping in the map-making
process. However, the $1/f$ noise can be considered negligible even
on the largest scales since the 5-year signal-to-noise level is so
high \citep{hinshaw_etal_2007}.

The extended temperature analysis mask (KQ75 which leaves approximately 72\% sky
usable coverage) is applied to the data
to eliminate regions contaminated by Galactic foreground and
point sources. Hemispherical masks of the Galactic North and
South (GN and GS), Ecliptic North and South (EN and ES) are also
applied together with the KQ75 mask for some parts of our analysis.
In this paper, we will refer to results
derived on the KQ75 sky-coverage as `full-sky' results for
convenience, to distinguish then from the corresponding
results on masked hemispheres.

\subsection{The two models}
\label{subsec_2model} In this section, we briefly review the
implementation of the two models discussed in our work. We use
$T'_{\rm{dmf}}$ and $T'_{\rm{acw}}$ to identify the simulated
realisations for each model, which are then combined with the beam
profiles and \WMAP\--like noise properties for each frequency band
concerned in the analysis. The realisations are performed at a
HEALPix resolution parameter $N_{\rm side} = 512$ with maximum
multipole, $\ell_{\rm max} = 1024$.

\subsubsection{Dipole modulation} \label{subsubsec_dipmod}

We follow \citet{eriksen_etal_2007}  and their implementation of
a multiplicative dipole modulation field as proposed in \citet{gordon_etal_2005}.

The CMB temperature field is defined as
\begin{equation}
    \tilde{T}(\vecn)=T(\vecn)[1+f(\vecn)], \label{eq_dm_Tfield}
\end{equation}
where $T(\vecn)=\sum_{\ell, m}a_{\ell m}Y_{\ell m}(\vecn)$ is a
statistically isotropic CMB temperature field with power
spectrum $C_{\ell}$ and $f(\vecn)$ is a
dipole modulation field with amplitude much less
than one but non-zero.

In their low $\ell$ analysis of the WMAP\ data, \citet{eriksen_etal_2007} have parameterised the
dmf model using four free
parameters, $q$ (normalisation amplitude), $n$ (tilt), $A$ (amplitude of dmf)
and $\vecp$ (direction of dmf), together with a fiducial Gaussian power-spectrum,
$C_{\ell}^{\rm fid}$, which was chosen to be the \WMAP\ best-fitting power spectrum.
The power spectrum and modulation field are
then written as
\begin{equation}
    C_{\ell; (q,n)}=q\left(\frac{\ell}{\ell_0}\right)^nC_{\ell}^{\rm fid} \
    {\rm{and}} \ f(\vecn; A,\vecp)=A\vecn \cdot \textbf{\textit{p}}.
\label{eq_dm_par}
\end{equation}
In fact, this formalism recognises that the best-fitting dmf may
also impact other cosmological parameters, but that over the limited
range in $\ell$ that the analysis is carried out, the practical
effect can be represented by a normalisation $q$ and spectral tilt
$n$. The posterior distributions of the parameters were then given
by an optimal Bayesian analysis.

In this work, we modulate Gaussian realisations of the CMB adopting
the most recent parameters estimated from the \WMAP5\ data in
\citet{hoftuft_etal_2009}. Their estimations seem to provide a
robust value of $\vecp$, whereas the magnitude $A$ seems to be a
function of $\ell_{\rm mod}$, the maximum value of $\ell$ included
in the analysis. We therefore simulate the dmf adopting the
best-fitting directions for the different $\ell_{\rm mod}$ ranges,
and consider values for the amplitude $A$ varying spanning the 68\%
confidence region for each range. These parameters sets are
summarised in Table~\ref{tab_dm_par}. Note that we apply these
parameters over the full range of $\ell$ used in the simulations.
Importantly, they find that $q$ and $n$ are consistent with unity
and zero respectively (for the V-band, the actual values are $q =
0.96 \pm 0.05$ and $n =  0.06 \pm 0.05$), thus we can simply assume
the \WMAP5\ best-fitting `lcdm+sz+lens' model power spectrum as our
fiducial spectrum, and that the presence of the dmf does not perturb
the best-fit cosmological parameters.

\begin{table}
\caption{The groups of parameters adopted for the dmf simulations.
The $(A,\vecp)$ values are estimated by \citet{hoftuft_etal_2009}
for different maximum values of multipole moment, $\ell_{\rm mod}$,
involved during the estimation. The
dipole axis $\vecp$ point towards $(l,b)$ in Galactic coordinates.}
\begin{center}
    \begin{tabular}{c|c|c|c}
        \toprule
        Band & $\ell_{\rm mod}$ & $A$ & $\vecp(l,b)$ \\
        \midrule
        \multirow{9}{*}{V} & \multirow{3}{*}{40} & $0.119-0.034$ &
        \multirow{3}{*}{$(224^{\circ},-22^{\circ})$} \\
        & & 0.119 & \\
        & & $0.119+0.034$ & \\
        \cmidrule(r){2-4}
        & \multirow{3}{*}{64} & $0.080-0.021$ & \multirow{3}{*}{$(232^{\circ},-22^{\circ})$} \\
        & & $0.080$ & \\
        & & $0.080+0.021$ & \\
        \cmidrule(r){2-4}
        & \multirow{3}{*}{80} & $0.070-0.019$ & \multirow{3}{*}{$(235^{\circ},-17^{\circ})$} \\
        & & $0.070$ & \\
        & & $0.070+0.019$ & \\
        \midrule
        \multirow{3}{*}{W} & \multirow{3}{*}{64} & $0.074-0.021$ &
        \multirow{3}{*}{$(232^{\circ},-22^{\circ})$} \\
        & & $0.074$ & \\
        & & $0.074+0.021$ & \\
        \bottomrule
    \end{tabular}
\end{center}
\label{tab_dm_par}
\end{table}

The simulated realisations $T'_{\rm dmf}(\vecn)$ are then given by
\begin{align}
\tilde{a}_{\ell m} & = \, b_{\ell}\sum_{\vecn}\tilde{T}(\vecn)Y_{\ell m}(\vecn), \\
T'_{\rm dm}(\vecn) & = \, \sum_{\ell, m}\tilde{a}_{\ell m}Y_{\ell
m}(\vecn) + N(\vecn).
\end{align}
These realisations are then convolved with the appropriate beam
transfer function  $b_{\ell}$, and noise realisations $N(\vecn)$ are
added. For each group of $(A,\vecp)$, we perform 2500 simulations
for comparison with observations.

\subsubsection{The ACW model} \label{subsubsec_acw}

Inflationary theory gives a firm prediction that the primordial
density perturbations should be isotropic without any preferred
direction. \citet{acw_2007} considered the breaking of this
rotational invariance by the explicit introduction of a preferred
direction, $\vecn'$, during the inflationary era thus modifying the
primordial power spectrum $P(k)$ to
\begin{equation}
P'(\veck; g(k),\vecn') = P(k)[1+g(k)(\veck \cdot \vecn'/k)^2],
\label{eq_prim_pwspc}
\end{equation}
where $\veck$ is the vector with magnitude $k$. $g(k)$ is a general function
representing the rotationally non-invariant part, which ACW have argued
can be approximated by a constant $g_{*}$.
The imprint of this preferred direction on the CMB is
then given by
\begin{equation}
S_{\ell m,\ell'm'} = \langle a_{\ell m}a^{*}_{\ell'm'}\rangle =
C_{\ell}\delta_{\ell \ell'}\delta_{mm'}+\Delta_{\ell m,\ell'm'},
\label{eq_covSPH}
\end{equation}
where $C_{\ell}$ is the CMB angular power spectrum and we define
\begin{equation}
\Delta_{\ell m,\ell'm'} = g_{*}\xi_{\ell m,\ell'm'}\int^{\infty}_{0}
dk \, k^2P(k)\Theta_{\ell}(k)\Theta_{\ell'}(k). \label{eq_Delta}
\end{equation}
$\xi_{\ell m,\ell'm'}$ are then geometric factors that describe
coupling between modes, $\ell$ to $\ell' = \ell, \ell + 2$ and $m$
to $m' = m, m\pm 1, m\pm 2$. The integral term is a generalisation
of the standard CMB power spectrum and may be computed using a
modified version of CAMB \citep{lewis_etal_2000}.

The contribution of $\Delta$ can contribute significantly
to the diagonal part of $\textbf{S}$ which affects the total angular power spectrum
and induces strong degeneracy between $g_{*}$ and the amplitude of
the power spectrum of scalar perturbations, $A_{s}$. However,
it can be ensured that a given choice of $g_{*}$ mainly affects the
anisotropic contribution by redefining the signal covariance matrix.
as
\begin{equation}
    S_{\ell m,\ell'm'} = \frac{C_{\ell}\delta_{\ell \ell'}\delta_{mm'}+\Delta_{\ell
    m,\ell'm'}}{1+g_{*}/3},
\end{equation}

In this work, we assume that the cosmological parameters are known
and given by the \WMAP5\ best-fitting `lcdm+sz+lens' model power
spectrum \citep{komatsu_etal_2009}. The two free parameters
remaining in the model, $(g_{*}, \vecn')$, have been estimated by
\citet{groe_hke_2009} from the \WMAP\ five-year data using Bayesian
analysis in an extended CMB Gibbs sampling framework. The estimation
was carried out over different $\ell$-ranges to investigate whether
the preferred direction $\vecn'$ is dependent on the $\ell$-range
selection. Indeed, a preferred direction and value for $g_{*}$ for
the ACW model was found for the V and W-bands at $2.5\sigma$ and
$3.8\sigma$ confidence-level, respectively (Table 1 of their paper),
although there is a $20^{\circ}$ deviation in Galactic longitude
between the two bands. For the range $\ell = [2,400]$ and KQ85
sky-coverage, the $g_{*}$ value is reported as $0.10\pm0.04$ for the
V-band and $0.15\pm0.039$ for W-band (95\% confidence), while for
the more conservative KQ75 sky-coverage the V-band $g_{*}$ value is
$0.10[-0.100,0.158]$. In \citet{hou_etal_2009}, we show that the
signal-to-noise-ratio (S/N) of the 5-year \WMAP\ temperature data
decreases to 1 at $\ell=399$ (385) on V- (W-) band, and the
additional smoothing imposed during data-processing suppresses
information on this scale by a further factor of 0.1 in spherical
harmonic space. Therefore, we choose different $g_{*}$ values
estimated from the $\ell$-range [2,400] within the 95\% confidence
region, to carry out the frequentist comparison. The parameters used
in this work are summarised in Table~\ref{tab_acw_par}.

\begin{table}
\caption{The groups of parameters used in our local extrema analysis
for ACW model. The parameters follow the estimation results of
\citet{groe_hke_2009} for $\ell$-range [2,400].}
\begin{center}
    \begin{tabular}{c|c|c|c}
        \toprule
        \multicolumn{2}{c}{V-band} & \multicolumn{2}{c}{W-band} \\
        $g_{*}$ & $\vecn' (l,b)$ & $g_{*}$ & $\vecn' (l,b)$ \\
        \midrule
        -0.100 & \multirow{5}{*}{$(130^{\circ},10^{\circ})$} & 0.111
        & \multirow{5}{*}{$(110^{\circ},10^{\circ})$} \\
        -0.050 & & 0.130 & \\
        0.050 & & 0.150 \\
        0.100 & & 0.170 \\
        0.158 & & 0.189 \\
        \midrule
        0.158 & $(110^{\circ},10^{\circ})$ & 0.189 &
        $(130^{\circ},10^{\circ})$ \\
        \bottomrule
    \end{tabular}
\end{center}
\label{tab_acw_par}
\end{table}

We then follow the algorithm of \citet{groe_hke_2009} to get the
simulated ACW realisations of the sky using
\begin{eqnarray}
T'_{\rm acw}(\vecn; g_{*}, \vecn') & = &
\sum_{\ell,m}b_{\ell}2^{\frac{\delta_{m0}-1}{2}}L_{\ell m}(g_{*},
\vecn')\eta_{\ell m}Y_{\ell m}(\vecn) \nonumber \\
 & & \mbox{} + N(\vecn),
\label{eq_acw_map}
\end{eqnarray}
where $L_{\ell m}$ %, as a precomputed kernel during simulations, is
is the Cholesky decomposition of the sparse covariance matrix
$\textbf{S}$, involving the best-fitting parameters $(g_{*},
\vecn')$ and cosmological model, and $\eta_{\ell m}$ is a set of
Gaussian random numbers with zero mean and unit variance. The factor
$2^{\frac{\delta_{m0}-1}{2}}$ is placed regarding the triangular
$L_{\ell m}$ and HEALPix data convention. Each simulation is
convolved with the appropriate \WMAP\ beams and noise realisations
are added. For each group of parameters, we perform 5000
simulations.

\subsection{Analysis and statistics}
\label{subsec_analysis}

In this section, we introduce the processing steps and statistics
utilised for the local extrema comparison between the observed
sky and simulations for our two selected models. Generally, the treatment
follows closely that of sections~2.5 and 2.6 of \citet{hou_etal_2009},
but is briefly reviewed here.

The data and simulations are treated in an identical way. Masks are
applied to the sky maps, then the best-fitting monopole and dipole
components for the region outside the mask are subtracted -- this is
a standard procedure in CMB data analysis. Since our previous
results on local extrema statistics suggest a close connection with
large angular scale modes on the sky, we also consider cases where
modes up to and including the quadrupole, $\ell \leq 5$ or $\ell
\leq 10$ are also subtracted before analysis. A modest smoothing is
then applied to the resulting sky maps where the angular scale of
smoothing is determined by a signal-to-noise normalisation
technique. In this work, we applied 43.485 arcmin and 45.064 arcmin
FWHM Gaussian beams to the masked sky maps of the V- and W-band
respectively. In order to be conservative and avoid potential
boundary effects in the analysis, the masks (for which all valid
pixels are set to unity) are also smoothed and only those pixels
with values larger than 0.90 are retained as valid. The local
extrema are then determined using the {\em hotspots} program in the
HEALPix package.

First, the one-point statistics (number, mean, variance, skewness
and kurtosis) of the local extrema temperature distribution are
computed for further hypothesis test. The hypothesis test
methodology introduced by \citet{LW05} is again adopted here. The
probability of finding a simulated statistic that falls below the
observed one is defined as $p$. As a two-sided test, we set the
significance level $\alpha = 0.05$ with the hypothesis
\begin{equation}
    H:p\in(\alpha/2,1-\alpha/2),
\end{equation}
 and perform the test twice to control
the Type \uppercase\expandafter{\romannumeral1} error ($\beta$ --
rejection of true hypothesis) and the Type
\uppercase\expandafter{\romannumeral2} error ($\gamma$ -- acceptance
of the false hypothesis) to be small.

Secondly, the two-point correlation functions of temperature pairs,
$\xi_{\rm TT}$, and spatial pair-counting, $\xi_{\rm PP}$, are
computed to make a further $\chi^2$ analysis. During the two-point
analysis, each realisation can be examined in a renormalised form,
$T(\vecn) = \nu(\vecn)\sigma_{\rm sky}$, where $\sigma_{\rm sky}$ is
the standard-deviation of the temperature field over the valid
region of \emph{each} realisation. We will continue to examine the
correlation dependence on the thresholds $\nu$,e and the large-scale
multipoles. For the $\xi_{\rm PP}$ computation, we utilise the
Hamilton estimator \citep{ham_1993} and use the same random sample
as in \citet{hou_etal_2009}. Since the $\{\xi_{\rm TT}(\theta)\}$
and $\{\xi_{\rm PP}(\theta)\}$ are asymmetrically distributed in a
given angular range, the $\chi^2$ statistics are optimised by
mapping the distributions to Gaussian ones $\{s(\theta)\}$,
\begin{equation}
\frac{\rm Rank \ of \ observed \ map}{\rm{Total \ number \ of \
maps}+1}=\frac{1}{\sqrt{2\pi}}\int_{-\infty}^{s}e^{-\frac{1}{2}t^2}\,dt.
\end{equation}
Then the probability of finding the number of simulations with a
$\chi^2$ value below the observed one can be determined.

\section{RESULTS AND DISCUSSIONS} \label{sec_results}

In this paper, we will not attempt to describe all the one- and two-point
results determined, but instead highlight some interesting cases that help
to build the conclusions.

Those results that are derived after subtraction
of the low--$\ell$ multipoles in the range $[0,\lrmv]$ are
denoted by $\lrmv$ for
simplicity, where $\lrmv = 1$, 2, 5 and 10.

In the frequentist comparison, we determine the probability $p$ described in
section~\ref{subsec_analysis} by counting the number of simulations
with statistical values  (either one-point statistics or the
associated $\chi^2$ measure for the two-point analysis)
below the observed one. For the one-point analysis, the
probabilities are subjected to a  hypothesis test twice, the first test
sets $\beta = \alpha$ and any rejections are listed in the tables and marked by
asterisks; the second test sets $\gamma = \alpha$, with the subsequent rejections
that were accepted by the first test marked by question-marks. If we
specify $\alpha = 0.05$, those cases rejected by the first test
indicate that the consistency of the observations with the model
concerned is rejected at the 95\% C.L., with an associated 5\%
probability of a Type \uppercase\expandafter{\romannumeral1} error
($\beta=0.05$). For those cases accepted by the second test, it
can then be asserted that such a consistency is accepted at the
0.05 significance level, with an associated 5\% probability of a Type
\uppercase\expandafter{\romannumeral2} error ($\gamma=0.05$).

\subsection{Dipole modulation results}
\label{subsec_dm}

A subset of the results are shown in
Table~\ref{tab_dm_1p_results}. We only present the local extrema mean
and variance statistics for
$\rm V^{A=0.119\pm0.034}_{\vecp=(224^\circ,-22^\circ)}$ and
$\rm W^{A=0.074\pm0.021}_{\vecp=(232^\circ,-22^\circ)}$ here, since the
number, skewness and kurtosis values are not revealing, and
the results from other groups of dmf parameters lead to similar
conclusions.

\begin{table*}
\caption{Frequencies of the extrema one-point statistics with lower values than the
\WMAP\ V and W-band five-year data. The dmf parameters are given as
superscripts and subscripts of the frequency label. The local maxima (minima)
results are denoted as `max' (`min'). NS, GN, GS, EN and ES
correspond to full-sky, Galactic North, Galactic South, Ecliptic
North and Ecliptic South sky-coverage outside of the KQ75 mask,
respectively. The values rejected by the hypothesis test are marked
by a * or ?. Values outside the $3\sigma$ confidence range are
underlined.}
\begin{center}
\scriptsize{
\begin{tabular}{rc|lllll|lllll}
    \toprule
    \multicolumn{2}{c}{\WMAP5, KQ75}
    & NS & GN & GS & EN & ES
    & NS & GN & GS & EN & ES\\
    \midrule
    \multicolumn{2}{c}{MEAN}
    & \multicolumn{5}{l}{$\lrmv=1$}
    & \multicolumn{5}{l}{$\lrmv=2$}\\

    \multirow{2}{*}{$\rm V^{A=0.085}_{(224^\circ,-22^\circ)}$} & max &
    0.1416 & 0.5252 & 0.0604 & 0.8584 & 0.0052* & 0.1436 & 0.8776 & 0.0600
    & 0.9628 & \underline{0.0000*} \\
    & min &
    0.0908 & 0.3908 & 0.0592 & 0.7980 & 0.0016* & 0.1304 & 0.3040 & 0.0596
    & 0.8720 & \underline{0.0008*} \\
    \multirow{2}{*}{$\rm V^{A=0.119}_{(224^\circ,-22^\circ)}$} & max &
    0.1216 & 0.5816 & 0.0392 & 0.8292 & 0.0064* & 0.1248 & 0.9088 & 0.0284?
    & 0.9468 & \underline{0.0000*} \\
    & min &
    0.0812 & 0.4460 & 0.0376 & 0.7632 & \underline{0.0012*} & 0.1108 & 0.3616 & 0.0324
    & 0.8408 & \underline{0.0012*} \\
    \multirow{2}{*}{$\rm V^{A=0.153}_{(224^\circ,-22^\circ)}$} & max &
    0.1312 & 0.8340 & 0.0060* & 0.9940* & \underline{0.0000*} & 0.1340 & 0.9856* & 0.0060*
    & \underline{0.9992*} & \underline{0.0000*} \\
    & min &
    0.0860 & 0.7536 & 0.0084* & 0.9892* & \underline{0.0000*} & 0.1180 & 0.6820 & 0.0068*
    & \underline{0.9988*} & \underline{0.0000*} \\

    & & \multicolumn{5}{l}{$\lrmv=5$}
    & \multicolumn{5}{l}{$\lrmv=10$}\\

    \multirow{2}{*}{$\rm V^{A=0.085}_{(224^\circ,-22^\circ)}$} & max &
    0.4056 & 0.9640 & 0.0164* & 0.9836* & \underline{0.0008*} & 0.4168
    & 0.9076 & 0.0260? & \underline{0.9988*} & \underline{0.0000*} \\
    & min &
    0.0596 & 0.3064 & 0.0328 & 0.9272 & \underline{0.0004*} & 0.0944
    & 0.6784 & 0.0396 & 0.9656 & \underline{0.0004*} \\
    \multirow{2}{*}{$\rm V^{A=0.119}_{(224^\circ,-22^\circ)}$} & max &
    0.3700 & 0.9748? & 0.0080* & 0.9760? & \underline{0.0008*} & 0.4072
    & 0.9684 & 0.0044* & \underline{1.0000*} & \underline{0.0000*} \\
    & min &
    0.0476 & 0.3648 & 0.0148* & 0.8968 & \underline{0.0000*} & 0.0916
    & 0.8520 & 0.0108* & 0.9984* & \underline{0.0000*} \\
    \multirow{2}{*}{$\rm V^{A=0.153}_{(224^\circ,-22^\circ)}$} & max &
    0.3900 & 0.9976* & \underline{0.0004*} & \underline{1.0000*} & \underline{0.0000*} & 0.4024
    & 0.9940* & \underline{0.0008*} & \underline{1.0000*} & \underline{0.0000*} \\
    & min &
    0.0532 & 0.7076 & 0.0020* & \underline{0.9992*} & \underline{0.0000*} & 0.0888
    & 0.9468 & 0.0024* & \underline{1.0000*} & \underline{0.0000*} \\
    \cmidrule(r){2-12}
    & & \multicolumn{5}{l}{$\lrmv=1$}
    & \multicolumn{5}{l}{$\lrmv=2$}\\

    \multirow{2}{*}{$\rm W^{A=0.053}_{(232^\circ,-22^\circ)}$} & max &
    0.1800 & 0.4380 & 0.1296 & 0.8904 & 0.0136* & 0.1528 & 0.7380 & 0.1036
    & 0.9556 & 0.0068* \\
    & min &
    0.0632 & 0.0860 & 0.1756 & 0.5716 & 0.0068* & 0.0704 & 0.0476 & 0.1984
    & 0.6148 & 0.0116 \\
    \multirow{2}{*}{$\rm W^{A=0.074}_{(232^\circ,-22^\circ)}$} & max &
    0.1760 & 0.5524 & 0.0852 & 0.9608 & 0.0028* & 0.1524 & 0.8244 & 0.0568
    & 0.9884* & \underline{0.0004*} \\
    & min &
    0.0624 & 0.1408 & 0.1076 & 0.7664 & \underline{0.0012*} & 0.0700 & 0.0816 & 0.1176
    & 0.8200 & 0.0024 \\
    \multirow{2}{*}{$\rm W^{A=0.095}_{(232^\circ,-22^\circ)}$} & max &
    0.1792 & 0.6564 & 0.0444 & 0.9888* & \underline{0.0004*} & 0.1492 & 0.8936 & 0.0288?
    & 0.9984* & \underline{0.0000*} \\
    & min &
    0.0624 & 0.2168 & 0.0616 & 0.8976 & \underline{0.0008*} & 0.0688 & 0.1392 & 0.0716
    & 0.9336 & \underline{0.0004*} \\

    & & \multicolumn{5}{l}{$\lrmv=5$}
    & \multicolumn{5}{l}{$\lrmv=10$}\\

    \multirow{2}{*}{$\rm W^{A=0.053}_{(232^\circ,-22^\circ)}$} & max &
    0.3256 & 0.8784 & 0.0304 & 0.9660 & \underline{0.0008*} & 0.4288
    & 0.7936 & 0.1144 & 0.9876* & 0.0076* \\
    & min &
    0.1444 & 0.0464 & 0.4172 & 0.3864 & 0.0532 & 0.0388
    & 0.1432 & 0.2716 & 0.4844 & 0.0076* \\
    \multirow{2}{*}{$\rm W^{A=0.074}_{(232^\circ,-22^\circ)}$} & max &
    0.3240 & 0.9344 & 0.0136* & 0.9916* & \underline{0.0004*} & 0.4224
    & 0.8800 & 0.0572 & 0.9980* & \underline{0.0008*} \\
    & min &
    0.1448 & 0.0952 & 0.2760 & 0.6568 & 0.0096* & 0.0376
    & 0.2368 & 0.1708 & 0.7528 & \underline{0.0008*} \\
    \multirow{2}{*}{$\rm W^{A=0.095}_{(232^\circ,-22^\circ)}$} & max &
    0.3192 & 0.9688 & 0.0044* & \underline{0.9992*} & \underline{0.0000*} & 0.4088
    & 0.9340 & 0.0248? & \underline{0.9996*} & \underline{0.0000*} \\
    & min &
    0.1380 & 0.1664 & 0.1756 & 0.8704 & \underline{0.0004*} & 0.0376
    & 0.3488 & 0.0924 & 0.9220 & \underline{0.0000*} \\
    \midrule

    \multicolumn{2}{c}{VARIANCE}
    & \multicolumn{5}{l}{$\lrmv=1$}
    & \multicolumn{5}{l}{$\lrmv=2$}\\
    \multirow{2}{*}{$\rm V^{A=0.085}_{(224^\circ,-22^\circ)}$} & max &
    0.0044* & 0.0080* & 0.2860 & 0.0128* & 0.1164 & 0.0420 & 0.0064* & 0.6112
    & 0.0544 & 0.3444 \\
    & min &
    0.0020* & 0.0040* & 0.2664 & 0.0120* & 0.0920 & 0.0272? & 0.0020* & 0.6544
    & 0.0576 & 0.2320 \\
    \multirow{2}{*}{$\rm V^{A=0.119}_{(224^\circ,-22^\circ)}$} & max &
    0.0028* & 0.0076* & 0.2412 & 0.0104* & 0.1024 & 0.0320 & 0.0068* & 0.5432
    & 0.0376 & 0.3196 \\
    & min &
    \underline{0.0008*} & 0.0040* & 0.2304 & 0.0096* & 0.0812 & 0.0164* & \underline{0.0008*} & 0.5988
    & 0.0392 & 0.2132 \\
    \multirow{2}{*}{$\rm V^{A=0.153}_{(224^\circ,-22^\circ)}$} & max &
    0.0020* & 0.0140* & 0.1664 & 0.0468 & 0.0300? & 0.0260? & 0.0136* & 0.4188
    & 0.1660 & 0.1076 \\
    & min &
    \underline{0.0008*} & 0.0068* & 0.1608 & 0.0472 & 0.0220? & 0.0144* & 0.0088* & 0.4668
    & 0.1684 & 0.0624 \\

    & & \multicolumn{5}{l}{$\lrmv=5$}
    & \multicolumn{5}{l}{$\lrmv=10$}\\

    \multirow{2}{*}{$\rm V^{A=0.085}_{(224^\circ,-22^\circ)}$} & max &
    0.0508 & 0.1860 & 0.3524 & 0.3336 & 0.0612 & 0.0616 & 0.5036
    & 0.0540 & 0.8244 & 0.0176* \\
    & min &
    0.0840 & 0.1100 & 0.2936 & 0.2292 & 0.0568 & 0.0240?
    & 0.3872 & 0.0360 & 0.6888 & 0.0084* \\
    \multirow{2}{*}{$\rm V^{A=0.119}_{(224^\circ,-22^\circ)}$} & max &
    0.0320 & 0.1720 & 0.2664 & 0.2448 & 0.0468 & 0.0380 & 0.5736
    & 0.0192* & 0.9404 & 0.0020* \\
    & min &
    0.0528 & 0.1000 & 0.2020 & 0.1608 & 0.0448 & 0.0148*
    & 0.4668 & 0.0136* & 0.8572 & \underline{0.0012*} \\
    \multirow{2}{*}{$\rm V^{A=0.153}_{(224^\circ,-22^\circ)}$} & max &
    0.0216? & 0.2772 & 0.1120 & 0.6692 & 0.0028* & 0.0208?
    & 0.6292 & 0.0064* & 0.9824* & \underline{0.0000*} \\
    & min &
    0.0392 & 0.1796 & 0.0888 & 0.5372 & 0.0028* & 0.0052*
    & 0.5212 & 0.0024* & 0.9408 & \underline{0.0000*} \\
    \cmidrule(r){2-12}
    & & \multicolumn{5}{l}{$\lrmv=1$}
    & \multicolumn{5}{l}{$\lrmv=2$}\\

    \multirow{2}{*}{$\rm W^{A=0.053}_{(232^\circ,-22^\circ)}$} & max &
    0.0044* & 0.0124* & 0.2716 & 0.0088* & 0.1576 & 0.0488 & 0.0064* & 0.5468
    & 0.0304 & 0.3528 \\
    & min &
    0.0064* & 0.0064* & 0.3660 & 0.0112* & 0.1532 & 0.0468 & 0.0108* & 0.7060
    & 0.0820 & 0.3124 \\
    \multirow{2}{*}{$\rm W^{A=0.074}_{(232^\circ,-22^\circ)}$} & max &
    0.0044* & 0.0156* & 0.2344 & 0.0140* & 0.1124 & 0.0456 & 0.0064* & 0.4884
    & 0.0516 & 0.2532 \\
    & min &
    0.0064* & 0.0080* & 0.3196 & 0.0232? & 0.1096 & 0.0416 & 0.0136* & 0.6636
    & 0.1180 & 0.2196 \\
    \multirow{2}{*}{$\rm W^{A=0.095}_{(232^\circ,-22^\circ)}$} & max &
    0.0044* & 0.0196* & 0.1996 & 0.0228? & 0.0764 & 0.0416 & 0.0100* & 0.4332
    & 0.0844 & 0.1796 \\
    & min &
    0.0052* & 0.0104* & 0.2812 & 0.0356 & 0.0776 & 0.0364 & 0.0184* & 0.6152
    & 0.1680 & 0.1572 \\

    & & \multicolumn{5}{l}{$\lrmv=5$}
    & \multicolumn{5}{l}{$\lrmv=10$}\\

    \multirow{2}{*}{$\rm W^{A=0.053}_{(232^\circ,-22^\circ)}$} & max &
    0.0468 & 0.1836 & 0.3376 & 0.1632 & 0.1352 & 0.1644 & 0.5216
    & 0.1048 & 0.5548 & 0.0588 \\
    & min &
    0.0636 & 0.1536 & 0.3024 & 0.2432 & 0.0544 & 0.0488 & 0.4416
    & 0.0744 & 0.6492 & 0.0216? \\
    \multirow{2}{*}{$\rm W^{A=0.074}_{(232^\circ,-22^\circ)}$} & max &
    0.0408 & 0.2152 & 0.2660 & 0.2556 & 0.0620 & 0.1452 & 0.5824
    & 0.0600 & 0.6948 & 0.0208? \\
    & min &
    0.0532 & 0.1876 & 0.2248 & 0.3564 & 0.0180* & 0.0440
    & 0.5076 & 0.0420 & 0.7928 & 0.0048* \\
    \multirow{2}{*}{$\rm W^{A=0.095}_{(232^\circ,-22^\circ)}$} & max &
    0.0320 & 0.2540 & 0.1904 & 0.3812 & 0.0260? & 0.1164
    & 0.6392 & 0.0380 & 0.8248 & 0.0040* \\
    & min &
    0.0424 & 0.2248 & 0.1672 & 0.4736 & 0.0060* & 0.0356
    & 0.5640 & 0.0244? & 0.8860 & \underline{0.0012*} \\

    \bottomrule
\end{tabular}
}
\end{center}
\label{tab_dm_1p_results}
\end{table*}

We first focus on the variance results from
Table~\ref{tab_dm_1p_results}. For the most general case, $\lrmv=1$,
the probabilities for both the V and W-bands in the GN and EN are
less-extreme than the unmodulated results as might be expected,
although there are still rejections of the model on the full-sky and
northern hemispheres. In addition, an improved agreement for
$\lrmv=2$ on EN and ES hemispheres indicates good consistency of the
real data with the dmf model which again should not be surprising
given its profile.  The results on hemispheres implies that an
increasing dmf amplitude may suppress the observed variance
anomalies. However, after subtraction of the first 10 multipoles,
new problems arise in the southern hemispheres, where the observed
variance of local extrema is now much lower than the model
prediction. This result is robust within the 68\% confidence region
of the dmf amplitudes tested here. It should also be noticed that,
contrary to the cases on hemispheres, the full-sky variance is
increasingly inconsistent with the model as the dmf amplitude
increases. This is suggestive that the dmf may not have the correct
profile over the full-sky in order to reconcile the variance
properties of the observed local extrema with the Gaussian
anisotropic simulations.

The observed mean statistics, which showed few anomalies when
compared to the Gaussian isotropic model, are not consistent with
the dmf, in particular for cases on the EN and ES. There are
$3\sigma$-level inconsistencies in both the V and W-bands that
demonstrate that the observed local extrema are not as extreme as
the model expectations on the ES, but are too extreme on the EN,
especially for $\lrmv=5$ and 10. Indeed, the discrepancies become
more significant with an increasing dmf amplitude. It is noteworthy
that the results for $\rm
V^{A=0.080\pm0.021}_{\vecp=(232^\circ,-22^\circ)}$ and $\rm
V^{A=0.070\pm0.019}_{\vecp=(235^\circ,-17^\circ)}$ (parameters
fitted for $\ell_{\rm mod}=64$ and 80, respectively) show weaker
evidence for a variance anomaly than found above, but the
inconsistency of the mean values remains at 3$\sigma$ C.L..
Therefore, we seem to have a `paradoxical' situation where a weaker
perturbed dmf is inadequate to explain the observed variance
difficulties whereas the mean results and full-sky variances argue
{\em against} a stronger dmf amplitude.

\begin{figure}
\begin{center}
    \includegraphics[width=0.48\textwidth,trim=-0.1cm 1.5cm 7.6cm 0.5cm]{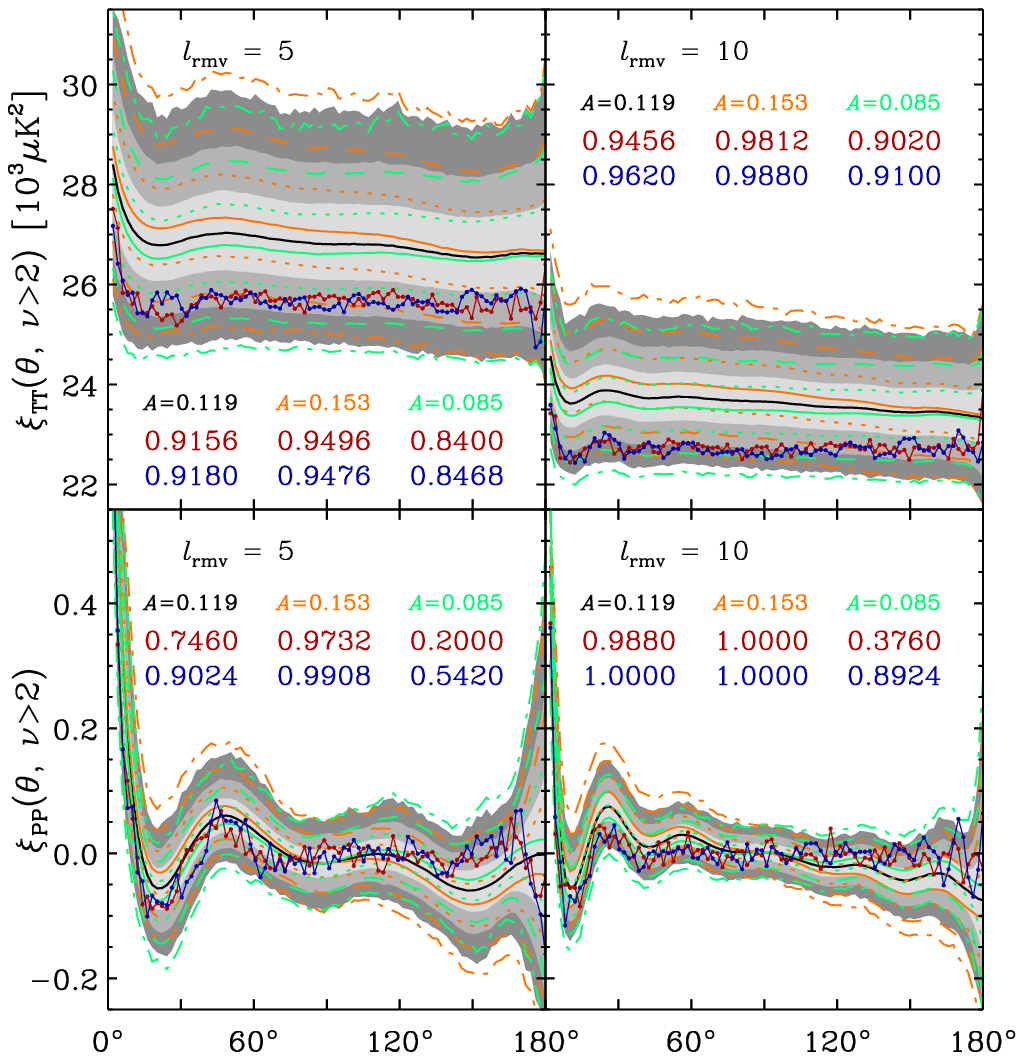}
\end{center}
\caption{The full-sky TT (upper panel) and PP (lower panel)
correlation functions (180 bins) of local extrema for $\nu>2$ with
$\lrmv=5$ and 10. The red (blue) lined-dots correspond to the
correlation function of local maxima (minima) observed in the \WMAP\
V-band data. The light, middle and dark gray shaded bands show,
respectively, the 68.26\%, 95.44\%, 99.74\% confidence regions
determined from 2500 MC simulations for dmf of $\rm
V^{A=0.119}_{\vecp=(224^\circ,-22^\circ)}$, and the black solid line
shows the median, as well as the confidence regions and the median
of $+/-1\sigma$ variation of the dmf amplitude shown by orange/green
lines (solid for the medians, dotted, dashed and dot-dashed for
boundaries of 68.26\%, 95.44\% and 99.74\% confidence regions,
respectively). The probabilities of finding a simulation with
corresponding $\chi^2$ values lower than the observed ones are
marked for both maxima (red) and minima (blue).}
\label{fig_dm2pcf_ns}
\end{figure}

The two-point results are more revealing. The full-sky results for
both $\xi_{\rm TT}(\nu\!<\!\infty)$ and $\xi_{\rm
PP}(\nu\!<\!\infty)$ show few deviations from the non-modulated
model and so cannot change the previous conclusions for these cases.
For $\xi_{\rm TT}(\nu\!>\!2)$, especially with $\lrmv=5$ and 10 as
shown by the upper panel of Figure~\ref{fig_dm2pcf_ns}, the
$\chi^2$-frequencies of observation are increasing as the amplitude
of the applied dmf increases. The relative spacing of the medians
and the confidence interval bands shows that such a shifting is
scale-dependent, decreasing on larger scales and providing evidence
that the observation does not favor a full-sky dmf since, as found
previously,  the shape of the observed $\xi_{\rm TT}(\nu\!>\!2)$
fits the non-modulated median quite well.

We then investigate $\xi_{\rm PP}(\nu\!>\!2)$ to find the impact on
the spatial distribution of the local extrema by the dmf. The lower
panel of Figure~\ref{fig_dm2pcf_ns} indicates that for realisations
with the first 5 or 10 multipoles subtracted, the local extrema tend
to be more clustered on scales less than $90^{\circ}$ while more
anti-clustered on larger scales, and this effect becomes more
pronounced for a stronger dmf amplitude. However, the full-sky
$\xi_{\rm PP}(\nu\!>\!2)$ medians in the equivalent unmodulated
cases fit the observations quite well, and simply oscillate about
the zero level on scales larger than $90^{\circ}$ indicating an
approximately uniform spatial distribution of local extrema with
$\nu\!>\!2$. Again, the empirical properties of the dmf are not
favoured by observations.

\begin{figure}
\begin{center}
    \includegraphics[angle=90,width=0.48\textwidth,trim=-0.5cm 12.8cm 0.4cm 0]{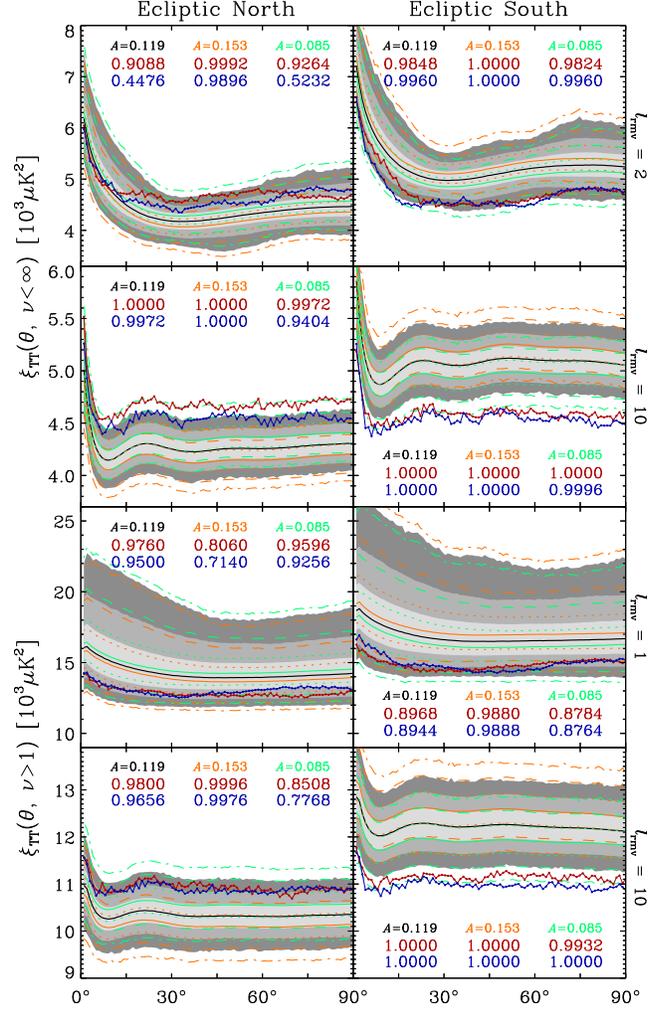}
\end{center}
\caption{The Ecliptic North and South $\xi_{\rm TT}(\nu\!<\!\infty)$
(the upper two panels for $\lrmv=2$ and 10) and $\xi_{\rm
TT}(\nu\!>\!1)$ (the lower two panels for $\lrmv=1$ and 10) with dmf
$\rm V^{A=0.119\pm0.034}_{\vecp=(224^\circ,-22^\circ)}$. The
nomenclature of the lined-dots, bands, lines and numbers follows the
same style as Figure~\ref{fig_dm2pcf_ns}.} \label{fig_dm2pcf_hem}
\end{figure}

The hemispherical results of $\xi_{\rm PP}$ show similar behavior to
the full-sky ones on scales less than $90^{\circ}$ and so we focus
on $\xi_{\rm TT}$ results for the Ecliptic hemispheres here. As
discussed in our previous work, on the GN and EN hemispheres with
$\lrmv=1$ and 2, the observed $\xi_{\rm TT}(\nu\!<\!\infty)$ is
suppressed on scales less than $20^{\circ}$ and $\xi_{\rm
TT}(\nu\!>\!1,\!2)$ are disfavoured at the $3\sigma$-level in
comparison with the Gaussian isotropic model. In contrast, the
results for the GS and ES, as well as for correlation functions with
$\lrmv=5$ and 10 over a variety of sky-coverages, show good
consistency with the predicted medians. The results presented in
Figure~\ref{fig_dm2pcf_hem} here indicate an offset in the expected
confidence regions between the north and south, and  the discrepancy
is enhanced with an increasing dmf amplitude. Considering the
profile of the dmf, the model practically attempts to solve the
suppression problem on the northern hemispheres by imposing an
effective north-south amplitude asymmetry that potentially improves
the fit in that part of the sky. On scales less than $15^{\circ}$,
the median of $\xi_{\rm TT}(\nu\!<\!\infty)$ with $\lrmv=2$ on EN
does match the observations well. However, the model behaves poorly
on the ES -- the observations lie below the lower $3\sigma$-level
for $\nu\!<\!\infty$ and below the $2\sigma$-level for $\nu\!>\!1$.
Moreover, after subtracting the first 5 or 10 multipoles, the
north-south asymmetry is increased so far that the model cannot fit
the observations, as quantified by the $\chi^2$-frequencies and
consistent with the behaviour of the one-point statistics. A simple
interpretation of these results would be that the profile of the dmf
is incorrect, and that the amplitude of the effect may also be a
function of $\ell$.

\subsection{ACW model results}
\label{subsec_acw}

\begin{figure}
\begin{center}
    \includegraphics[width=0.48\textwidth,trim=0 0 4.6cm 0]{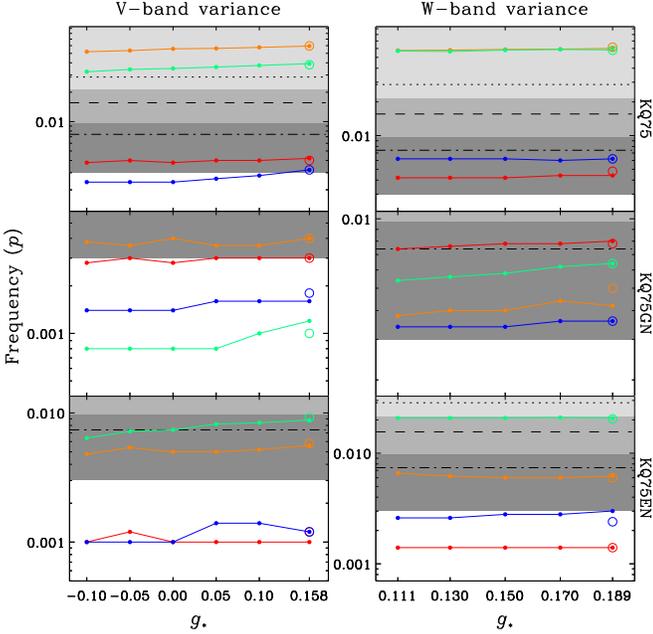}
\end{center}
\caption{The frequencies ($p$) for which the local extrema variances
of 5000 ACW simulations fall below the \WMAP5 V and
W-band observations as a function of $g_{*}$. Results are shown for the
the full-sky (upper panel), GN (middle) and EN (lower).
The red (blue) dots and lines correspond to the maxima
(minima) frequencies for $\lrmv=1$, whilst the orange (green)
values correspond to $\lrmv=2$. The centres of the coloured circles
correspond to the cases summarised in the last row of
Table~\ref{tab_acw_par}. Note that the red and blue circle in
the lower-left panel overlap. The light, middle and dark grey-shaded bands
fix the rejection regions in our two-sided hypothesis test with
confidence level 95.0\%, 97.5\% and 99.0\%, respectively, with the
Type \uppercase\expandafter{\romannumeral1} error $\beta \equiv
\alpha$. The dotted, dashed and dot-dashed lines fix the lower-limit
of the acceptance regions with $\alpha=0.05$, 0.025 and 0.01,
respectively, with the Type \uppercase\expandafter{\romannumeral2}
error $\gamma \equiv \alpha$. A logarithmic coordinate is applied
to the vertical axis to clearly show the acceptance/rejection
boundaries.} \label{fig_acw_var}
\end{figure}

Generally speaking, the local extrema results in the ACW scenario show few
differences from the corresponding isotropic results in \citet{hou_etal_2009}.
For the northern hemispheres, the
observed variances are still rejected at 95\% confidence level for
$\lrmv=1$ and 2, while other one-point statistics are quite
consistent with the model expectations, as judged by the hypothesis
tests. Improvements in consistency occur again for $\lrmv=5$ and 10.
Nevertheless, we further investigate the variance results
on the full-sky, GN and EN as a function of $g_{*}$ (Figure~\ref{fig_acw_var}),
and more sophisticated acceptance/rejection boundaries are
applied by setting different significance levels -- $\alpha = 0.050$,
0.025 and 0.010.

The $p$-values for the full-sky V-band variances exhibit an increasing trend
$g_{*}$ value (note that we have added our previous
results for isotropic ($g_{*}=0.00$) simulations to the plot), although
such a trend is not apparent for the W-band. The frequencies of the
$\lrmv=1$ minima (blue lines and dots in Figure~\ref{fig_acw_var}) are
asserted to be rejected at 99\% C.L., while not being 99\% rejected
(but still not accepted at the same level) when $g_{*}$ increases to
its 95\% confidence upper limit, 0.158. Despite the apparent
increasing trend in the V-band, the $\lrmv=1$ maxima (red lines and dots)
variances are still not accepted at 99\% C.L., while results of
$\lrmv=2$ are accepted within 95\% C.L., also showing the increasing
trend. This indicates that the ACW isotropy violation model
shows mild improvement in the agreement with observations
of the full-sky extrema variances.
Nevertheless, the improvement for the
cases with $\lrmv=1$ is inadequate to be accepted by the hypothesis test
even at a significance level of 0.01.
The results on GN and EN show similar features. The observed
maxima variance on the W-band GN hemisphere increases with
$g_{*}$ and is accepted at 0.01 significance level, as are the minima on
the V-band EN hemisphere with $\lrmv=2$. There are no other
acceptance/rejection changes as a function of $g_{*}$.

It is a natural concern as to the extent to which an incorrectly
adopted value for the preferred direction can affect results.
\citet{groe_hke_2009} found a $20^{\circ}$ deviation of the
best-fitting $\vecn'$  between V and W-band. Therefore, we assume
that such a deviation represents the typical estimation error on the
preferred direction and test its impact by, for example, fixing the
maximum $g_{*}$ estimated from the W-band but setting $\vecn'$ as
that estimated from the V-band, as summarised in the last row of
Table~\ref{tab_acw_par}. However, as indicated by the coloured
circles in Figure~\ref{fig_acw_var}, no qualitative change is seen.

\begin{figure}
\begin{center}
    \includegraphics[angle=90,width=0.48\textwidth,trim=-0.5cm 12.8cm 0.4cm 0]{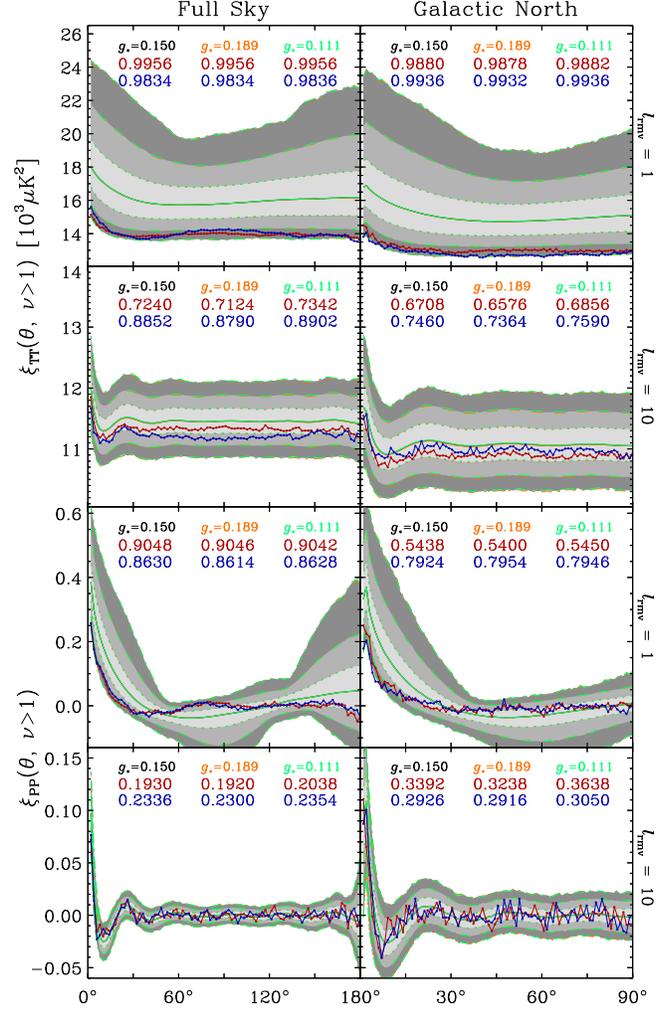}
\end{center}
\caption{The full-sky and Galactic North $\xi_{\rm TT}(\nu\!>1)$
(the upper two panels for $\lrmv=1$ and 10) and $\xi_{\rm
PP}(\nu\!>\!1)$ (the lower two panels for $\lrmv=1$ and 10) with
5000 ACW simulations on W-band, $g_{*}=0.150\pm0.039$,
$\vecn'=(130^\circ,10^\circ)$. The nomenclature of the lined-dots,
bands, lines and numbers follows the same style as
Figure~\ref{fig_dm2pcf_ns}. Note that the medians and the
boundaries of confidence region are more or less overlapped.}
\label{fig_acw2pcf}
\end{figure}

Some of the two-point correlation results of ACW simulations are
highlighted in Figure~\ref{fig_acw2pcf}. Even though the
$\chi^2$-frequencies decrease for larger $g_{*}$ values, the
$3\sigma$-level suppression of $\xi_{\rm TT}(\nu\!>1, \lrmv=1,2)$ on the
full-sky, GN and EN is still robust for the W-band, and the ACW model
does not cause the $\chi^2$-frequencies to improve sufficiently until
the first 5 or 10 multipoles are subtracted. The confidence regions
deviate little as $g_{*}$ varies within its 95\% estimation error,
constraining the observations almost at the same level as the
isotropic model, for both $\xi_{\rm TT}$ and $\xi_{\rm PP}$. It is
thus convincingly indicated that the ACW model alone is unable
to reconcile the local extrema anomalies originally found in the
context of an isotropic Gaussian model of CMB fluctuations.

\section{CONCLUSIONS} \label{sec_conclusion}

In this paper, the statistical properties of local temperature
extrema in the 5-year \WMAP\ data have been compared with two models
that break rotational invariance -- dipole modulation and the ACW
model. Such a comparison was motivated by the fact that our previous
analysis of local extrema statistics \citep{hou_etal_2009}
demonstrated an unlikely hemispherical asymmetry in the GN and EN as
a consequence of an extremely low-variance compared to the
expectations of a Gaussian isotropic scenario. Moreover, the 5-year
\WMAP\ data indicate significant detections of these models on the
basis of power spectrum analyses.

We employ Gaussian MC simulations encoding the features of these two
models to establish the statistical basis for testing whether the
breaking of rotational invariance by the models can afford a
satisfactory explanation of local extrema anomalies. As in previous
work, both one and two-point statistics, as well as their dependence
on large angular-scale modes have been studied. Both models are
parameterised by a set of amplitudes and preferred directions as
established by independent Bayesian analyses in
\citet{hoftuft_etal_2009} and \citet{groe_hke_2009}. Our analysis
has been carried out by sampling the former within their 95\%
confidence intervals, whilst imposing the fixed preferred direction
found for each band, since the latter are estimated to be robust for
each model. The V and W-band data are considered here.

In fact, the local extrema studies based on one and two-point
statistics indicate that neither model provides a satisfactory
solution to observed properties.
In particular, the ACW model is probably the least interesting
in this context -- both the one-point analysis and two-point studies
show similar features to the isotropic results, including
their $\ell$-dependence. The dipolar modulation field, however,
may itself be constrained by our analysis.

Results determined from division of the data into
hemispheres implies that a dmf with significant amplitude may
suppress the observed variance anomalies.
In particular, the $\lrmv=2$ values on EN and ES hemispheres
indicates consistency of the data with the model. However, after
subtraction of the first 10 multipoles, new problems arise in the
southern hemispheres, where the observed variance of local extrema is
now much lower than the model prediction.
This is suggestive that the dmf
may not have the correct profile over the full-sky
in order to reconcile the variance properties of the observed local extrema
with the Gaussian anisotropic simulations, and that the amplitude of the effect may also be a function
of $\ell$. Moreover,
the observed mean statistics, which showed few anomalies when compared
to the Gaussian isotropic model, are not consistent and contradict the
variance results by requiring a weaker amplitude.

Our analysis indicates that neither a simple dipole modulated field
model nor a rotational-invariance breaking model can satisfactorily
explain the local extrema anomalies present in the \WMAP\ data. On the one
side, the rotational-invariance model affects the large-scale
temperature amplitudes too little to significantly affect the local
extrema statistics. On the other side, a modulation type model needs a
more elaborate spatial structure than a simple dipole to fully fit the
data. These issues remain interesting for further investigation.

\section*{ACKNOWLEDGEMENTS}
Z.H. acknowledges the support by Max-Planck-Gesellschaft Chinese
Academy of Sciences Joint Doctoral Promotion Programme
(MPG-CAS-DPP). Most of the computations were performed at the
Rechenzentrum Garching (RZG) of Max-Planck-Gesellschaft and the IPP.
Some of the results in this paper have been derived using the
HEALPix \citep{gorski_etal_2005} software and analysis package. We
acknowledge use of the Legacy Archive for Microwave Background Data
Analysis (LAMBDA) supported by the NASA Office of Space Science.

\label{lastpage}
\end{document}